\documentclass[aps,prb,showpacs,reprint,groupaddress]{revtex4-1}

\usepackage{graphicx}
\usepackage{epstopdf}
\usepackage{hyperref}
\usepackage{here}

\begin{document}


\title{Intrinsic pinning and the critical current scaling of clean epitaxial Fe(Se,Te) thin films}


\author{Kazumasa Iida}
\email[Electronical address:\,]{k.iida@ifw-dresden.de}
\author{Jens H\"{a}nisch}
\author{Elke Reich}
\author{Fritz Kurth}
\author{Ruben H\"{u}hne}
\author{Ludwig Schultz}
\author{Bernhard Holzapfel}
\affiliation{Institute for Metallic Materials, IFW Dresden, 01171 Dresden, Germany}
\author{Ataru Ichinose}
\author{Masafumi Hanawa}
\author{Ichiro Tsukada}
\affiliation{Central Research Institute of Electric Power Industry, 2-6-1 Nagasaka, Yokosuka, Kanagawa 240-0196, Japan}
\author{Michael Schulze}
\author{Saicharan Aswartham}
\affiliation{Institute for Solid State Research, IFW Dresden, 01171 Dresden, Germany}
\author{Sabine Wurmehl}
\author{Bernd B\"{u}chner}
\affiliation{Institute f\"{u}r Festk\"{o}perphysik Technische Universit\"{a}t Dresden, 01062 Dresden, Germany}
\affiliation{Institute for Solid State Research, IFW Dresden, 01171 Dresden, Germany}

\date{\today}

\begin{abstract}
We report on the transport properties of clean, epitaxial Fe(Se,Te) thin films prepared on Fe-buffered MgO (001) single crystalline substrates by pulsed laser deposition. Near $T_{\rm c}$ a steep slope of the upper critical field for $H||ab$ was observed (74.1\,T/K), leading to a very short out-of-plane coherence length, $\xi_{\rm c}(0)$, of 0.2\,nm, yielding 2$\xi_{\rm c}(0)\approx0.4\,{\rm nm}$. This value is shorter than the interlayer distance (0.605\,nm) between Fe--Se(Te) planes, indicative of modulation of the superconducting order parameter along the $c$-axis. An inverse correlation between the power law exponent $N$ of the electric field--current density($E$--$J$) curve and the critical current density, $J_{\rm c}$, has been observed at 4\,K, when the orientation of $H$ was close to the $ab$-plane. These results prove the presence of intrinsic pinning in Fe(Se,Te). A successful scaling of the angular dependent $J_{\rm c}$ and the corresponding exponent $N$ can be realized by the anisotropic Ginzburg Landau approach with appropriate $\Gamma$ values 2$\sim$3.5. The temperature dependence of $\Gamma$ behaves almost identically to that of the penetration depth anisotropy.
\end{abstract}


\maketitle

\section{Introduction}
Investigating upper critical field ($H_{\rm c2}$) and its anisotropy ($\Gamma_{H_{\rm c2}}$) has been always a primal and common practice, since these values directly or indirectly yield important physical parameters, e.g. coherence length and mass anisotropy, $\Gamma_{\xi}= \xi_{\rm ab}/\xi_{\rm c}=\sqrt{m_{\rm c}/m_{\rm ab}}$, and penetration depth anisotropy, $\Gamma_{\lambda}= \lambda_{\rm c}/\lambda_{\rm ab}=\xi_{\rm ab}/\xi_{\rm c}$, in the case of single band superconductors, where $ab$ and $c$ are the crystallographic directions.

Among the Fe-based superconductors, Fe(Se,Te) single crystals show the steepest slope of $H_{\rm c2}$ ($|dH_{\rm c2}/dT|$=26\,T/K) for $H\parallel ab$ near $T_{\rm c}$.\cite{02} The evaluated out-of-plane coherence length, $\xi_{\rm c}$, was 0.35\,nm, which is shorter than the interlayer distance between Fe--Se(Te) planes, strongly indicative of the presence of intrinsic pinning. Additionally, a large $H_{\rm c2}$ is also highly expected at low temperatures, necessitating high magnetic fields to explore the magnetic phase diagram.

Recently we have applied the anisotropic Ginzburg Landau (AGL) scaling\cite{05} to the angular dependent critical current density, $J_{\rm c}(\Theta$), measured on epitaxial Co-doped BaFe$_{2}$As$_{2}$ (Ba-122) thin films,\cite{06,08} albeit this theory has been developed for single-band superconductors. Nevertheless the scaling parameters, $\Gamma$, have a temperature dependence and follow $\Gamma_{H_{\rm c2}}$.\cite{07} We have also found that the AGL approach is applicable to epitaxially grown LaFeAs(O,F) (La-1111) thin films.\cite{09} These results indicate that this approach may be also valid for evaluating $\Gamma_{H_{\rm c2}}$ even for other Fe-based superconducting materials regardless of their multi--band structures. Most importantly, this approach does not require a high-field magnet.

Scaling of the angular dependent resistivity for a NdFeAs(O,F) single crystal by the AGL approach has been reported by Jia $et$ $al$.\cite{10} They concluded that the AGL scaling can be applied in the Fe-based system since the anisotropies from different bands are quite close to each other.

Another method to evaluate $\Gamma_{\xi}$ through  $J_{\rm c}$ measurements has been reported by Ko{\'n}czykowski $et$ $al$.\cite{11} They have measured the critical current densities on LiFeAs single crystals along their principal directions namely $j_{ab}$ and $j_c$ with fields applied to the $ab$-plane. The ratio $j_{ab}/j_c$ directly yields $\Gamma_{\xi}$ in the strong pinning regime. Later, van der Beek $et$ $al$. have pointed out through their phenomenological approach that the field-angular dependence of critical current density ($J_{\rm c}(H, \Theta$)) for multi-band superconductors with a relatively large coherence length anisotropy and/or small point-like pinning centers behave similar to that of single-band superconductors.\cite{12}

Both arguments (i.e. Jia $et$ $al$.\cite{10} and van der Beek $et$ $al$.\cite{12}) seem to justify the implementation of the AGL scaling to other multi-band superconducting systems like Fe(Se,Te) as long as the above mentioned condition is held. Hence, it is obvious to apply the AGL scaling to clean, epitaxial Fe(Se,Te) films. 

Epitaxial Fe(Se,Te) thin films have been fabricated via pulsed laser deposition, PLD, by several groups.\cite{13,14,15,16} Recently the $J_{\rm c}(H, \Theta$) measurements on Fe(Se,Te) films have been reported by Bellingeri $et$ $al$.\cite{17} They have observed $c$-axis correlated defects in Fe(Se,Te) films on SrTiO$_3$ (001) substrates by scanning tunneling microscope, which led to enormous $J_{\rm c}$ peaks at $H\parallel c$. Similar $c$-axis peaks in $J_{\rm c}$ have been reported in Fe(Se,Te) films on CaF$_2$ (001) substrates by Mele $et$ $al$.\cite{33} In contrast, no correlated defects are observed in Fe(Se,Te) films on LaAlO$_3$ (001).\cite{17} We have also fabricated epitaxial Fe(Se,Te) films on Fe-buffered MgO (001) substrates with sharp out-of- and in-plane texture.\cite{18} The film showed no $c$-axis peak in $J_{\rm c}(\Theta$) measurements indicative of the absence of correlated defects in the film.

In this paper, we present various transport measurements for epitaxial Fe(Se,Te) thin films grown on Fe-buffered MgO (001) and discuss possible intrinsic pinning followed by the AGL scaling behavior. The evaluated anisotropy by $J_{\rm c}$ scaling is observed to increase with decreasing temperature, which is different from what we observed in Co-doped Ba-122 and La-1111.\cite{06,07,08,09}

\section{Experiments}
Fe(Se,Te) films have been deposited on Fe-buffered MgO (001) single crystalline substrates at 450\,$^\circ$C by ablating an Fe(Se,Te) single-crystal target with a KrF excimer laser in an ultra high vacuum chamber.

The PLD target was prepared by a modified Bridgman technique yielding an Fe(Se,Te) crystal with the nominal composition of Fe:Se:Te=1:0.5:0.5. For the target growth, stoichiometric amounts of pre-purified metals were sealed in an evacuated quartz tube. The tube was placed in a horizontal tube furnace and heated up to 650\,$^\circ$C and kept at that temperature for 24\,h. The furnace was then heated to 950\,$^\circ$C and the temperature was kept constant for 48\,h. Finally, the furnace was cooled down with a rate of 5\,$^\circ$C/h to 770\,$^\circ$C, followed by furnace cooling. We yield crystals with dimensions up to cm-size. A bulk $T_{\rm c}$ of 13.6\,K was recorded by a superconducting quantum interference device magnetometer.

The optimum deposition temperature, $T_{\rm s}$, is 450\,$^\circ$C since further increase or decrease in $T_{\rm s}$ leads to a slight decrease of $T_{\rm c}$. This optimum $T_{\rm s}$ is also  in good agreement with Ref.\cite{14}. A laser repetition rate of 7\,Hz was employed. A base pressure of 10$^{-10}$\,mbar is maintained. This low pressure level is increased to 10$^{-8}$\,mbar during the deposition due to degassing. Prior to the deposition, an Fe buffer layer was prepared at room temperature with a laser repetition rate of 5\,Hz, followed by a high-temperature annealing at 750\,$^\circ$C for 20 minutes. $In$-$situ$ reflection high-energy electron diffraction showed only streak patterns for all films, proving a flat surface of the Fe(Se,Te) film.

The films were structurally characterized by means of X-ray diffraction in $\theta/2\theta$ scans at Bragg-Brentano geometry with Co-K$_\alpha$ radiation and a texture goniometer system operating with Cu-K$_\alpha$ radiation.

A gold cap layer was deposited on the films at room temperature by PLD to prevent it from any damage during sample preparation and to achieve low contact resistance.

For transport measurements, 3 bridges namely "Bridge 1, 2, and 3" of 0.25--0.5\,mm width and 1\,mm length were fabricated from different sample areas by ion beam etching. Silver paint was employed for electrical contacts. $I$-$V$ characteristics on these samples were measured with four-probe configuration by a commercial Physical Property Measurement System (PPMS, Quantum Design). A voltage criterion of 1\,$\rm\mu Vcm^{-1}$ was employed for evaluating $J_{\rm c}$. In the angular-dependent $J_{\rm c}$ measurements, the magnetic field, $H$, was applied in  maximum Lorentz force configuration ($H$ perpendicular to $J$, where $J$ is current density) at an angle $\Theta$ from the $c$-axis.

\section{Results and discussion}
\begin{figure}
	\centering
		\includegraphics[width=\columnwidth]{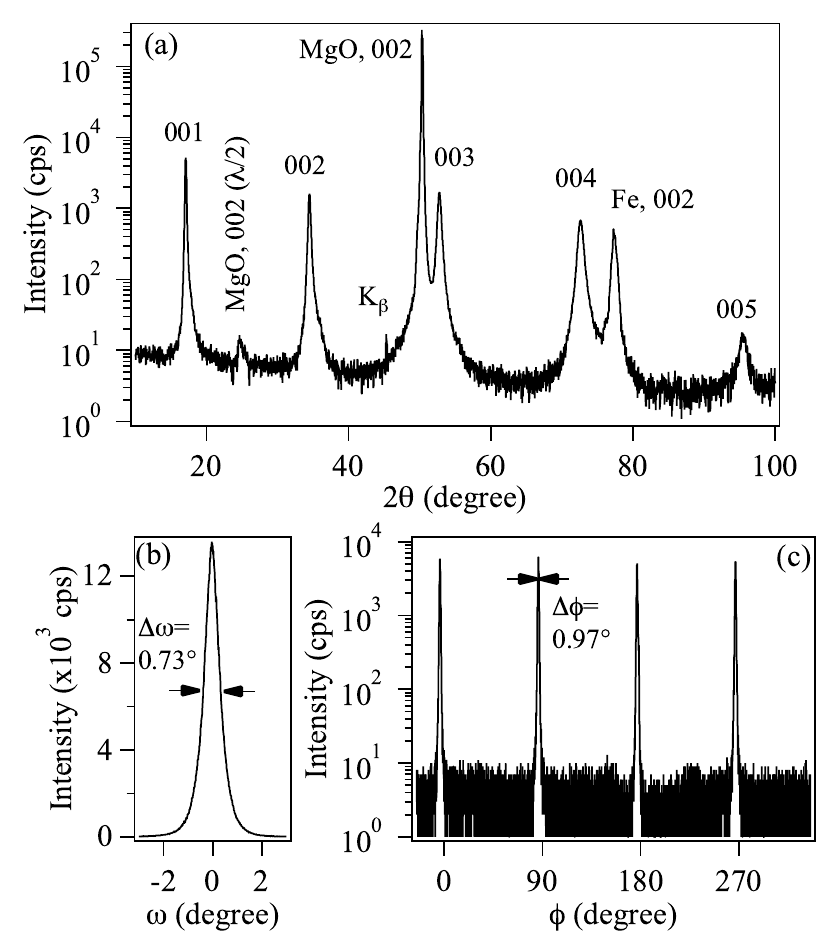}
		\caption{(a) $\theta/2\theta$ scan of Fe(Se,Te) on Fe-buffered MgO (001) substrate. (b) The rocking curve of the 001 reflection shows a $\Delta\omega$ of 0.73$^{\circ}$. (c) The 101 reflection of the $\phi$ scan of Fe(Se,Te) exhibits a four-fold symmetry. The average $\Delta\phi$ is 0.97$^\circ$.} 
\label{fig:figure1}
\end{figure}

\begin{figure*}
	\centering
		\includegraphics[width=14cm]{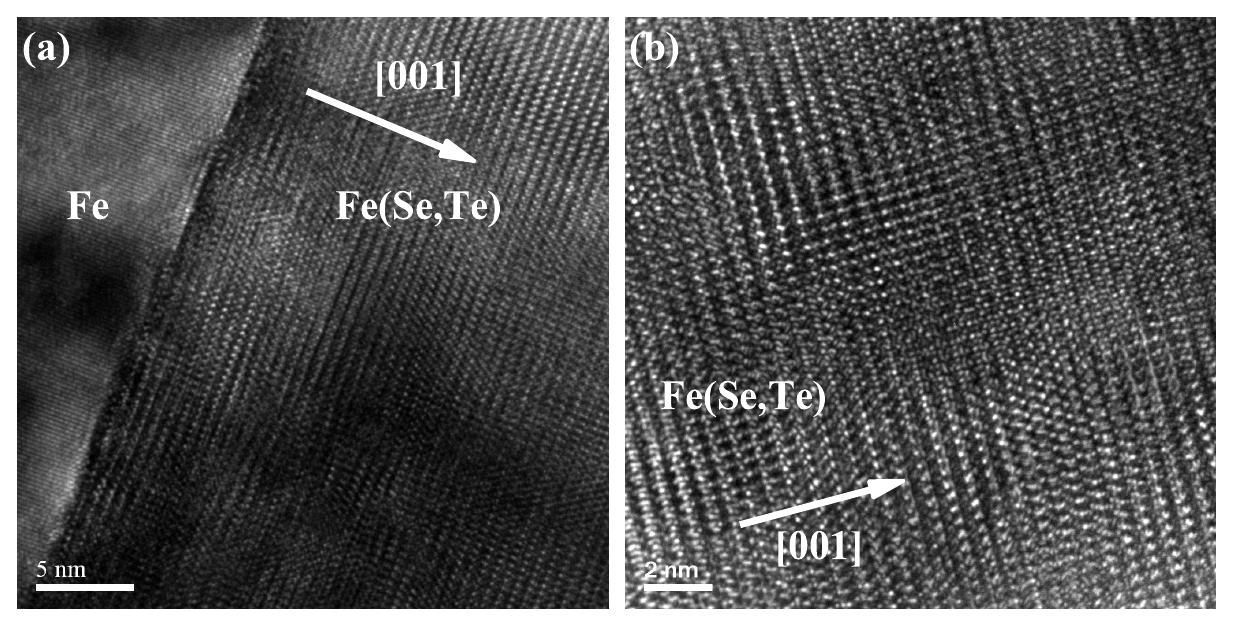}
		\caption{(Color online) (a) Cross-sectional bright-field TEM image of the Fe(Se,Te) thin film grown on Fe-buffered MgO (001) in the vicinity of interface. No crystallographic disordering is observed at the interface between Fe(Se,Te) and Fe layers.  Fe(Se,Te) layers contained no extended defects, however, small angle grain boundaries were observed. (b) High resolution TEM micrograph of the Fe(Se,Te) thin film.} 
\label{fig:figure2}
\end{figure*}

Structural characterization of Fe(Se,Te) films by means of X-ray diffraction is summarized in Figs.\,\ref{fig:figure1}. $\theta/2\theta$ scans confirmed that the Fe(Se,Te) layer was grown in $c$-axis textured (i.e. [001] perpendicular to the substrate) with high phase purity (Fig.\,\ref{fig:figure1}(a)). The rocking curve of the 001 reflection showed a full width at half maximum ($\Delta\omega$) of 0.73$^{\circ}$, which proves a good out-of-plane texture (Fig.\,\ref{fig:figure1}(b)). The 101 pole figure measurements ($\Psi=58.8^\circ$ and $2\theta=28.1^\circ$, not shown in this paper) and the corresponding $\phi$ scan of the Fe(Se,Te) film exhibited a clear four-fold symmetry and an average full width at half maximum ($\Delta\phi$) of 0.97$^\circ$ (Fig.\,\ref{fig:figure1}(c)). These results are evident that the film was epitaxially grown and of high crystalline quality. Here the epitaxial relationship between the Fe(Se,Te) layer, the Fe buffer layer, and the MgO substrate is (001)[100]Fe(Se,Te)$\|$(001)[110]Fe$\|$(001)[100]MgO. 

\begin{figure*}
	\centering
			\includegraphics[width=17cm]{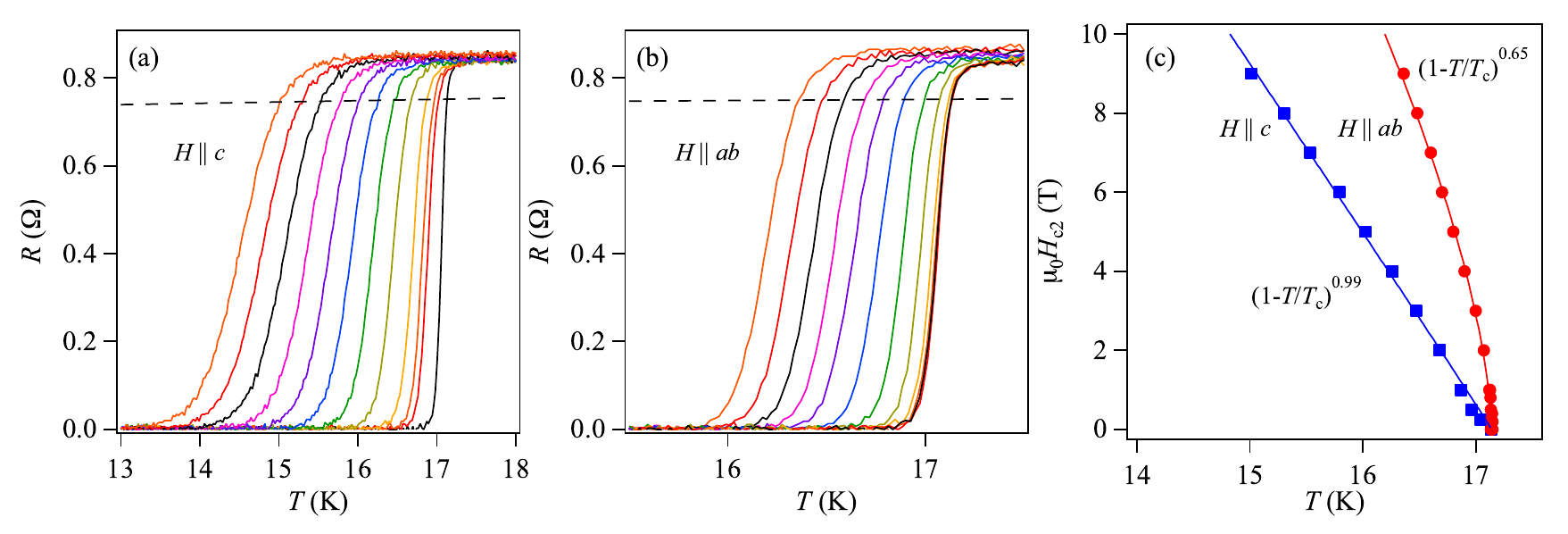}
		\caption{(Color online) Resistance traces of Fe(Se,Te) films measured in applied fields up to 9\,T for (a) $H\|c$ and (b) $H\|ab$. The broken lines indicate a 90\% of normal state resistance at 20\,K. (c) The $\mu_0H_{\rm c2}(T)$ for both major directions (solid circle: $H\|ab$ and solid square: $H\|c$). The solid red and black lines are fits using the $(1-T/T_{\rm c})^n$.} 
\label{fig:figure3}
\end{figure*}

Shown in Figs.\,\ref{fig:figure2} is the cross-sectional TEM image for an Fe(Se,Te) thin film in the vicinity of the interface. The respective layer thicknesses of Fe buffer and Fe(Se,Te) film are confirmed to be 18\,nm and 75\,nm. It is further obvious that a sharp interface between Fe(Se,Te) and Fe layer is realized, which is similar to the Ba-122/Fe bilayer system.\cite{35} Additionally, Fe(Se,Te) layers contained neither extended defects nor large angle grain boundaries. However, a small density of dislocations and small angle grain boundaries are observed.

The superconducting transition temperature, $T_{\rm c}$, of "Bridge\,3", which is defined as 90\% of normal resistance at 20\,K, is 17.3\,K under zero magnetic field (Fig.\,\ref{fig:figure3}\,(a), (b)). This $T_{\rm c}$ value is higher than the bulk value presumably due to compressive strain.\cite{19} Some bridges (e.g. Bridge\,1 and 2) including an un-patterned film were also measured, and all traces show almost the same $T_{\rm c}$ value with a variation of 0.1\,K. Additionally, the field dependencies of $J_{\rm c}$ for all bridges are almost identical, indicative of a homogeneous film (see Fig.\,\ref{fig:figureS1} in the Appendix A).

When magnetic fields are applied to the film, an apparent shift of $T_{\rm c}$ to lower temperatures is observed for both crystallographic directions. This shift together with a broadening of the transition is more significant for $H$\,$\parallel$\,${c}$ than for $H$\,$\parallel$\,${ab}$, which is typical for Fe-based superconductors with high Ginzburg numbers. Fig.\,\ref{fig:figure3}\,(c) shows the temperature dependence of $H_{\rm c2}$ for field parallel and perpendicular to the $c$-axis. For both directions, $H_{\rm c2}$ is proportional to $(1-T/T_{\rm c})^n$ and the respective exponents $n$ for $H\|c$ and $H\|ab$ are 0.99 and 0.65. The exponent $n=0.65$ for $H\|ab$ is close to 0.5, which is expected for layered compounds.\cite{34}

Near $T_{\rm c}$, slopes of $\vert \frac{d\mu_{0}H^{||c}_{c2}}{dT} \vert_{T_{c}} = 4.4$\,T/K and $\vert \frac{d\mu_{0}H^{||ab}_{c2}}{dT} \vert_{T_{c}} = 74.1$\,T/K were recorded, resulting in the anisotropy of orbital upper critical field, $\Gamma_{H^{\rm orb.}_{\rm c2}}(0)=16.8$ through the conventional Werthamer-Helfand-Hohenberg (WHH) theory.\cite{38} Such an extremely steep slope for $H\|ab$ has been also observed in strained Fe(Se,Te) films, indicating a very short out-of-plane coherence length.\cite{20} Here the out-of-plane coherence length at low temperatures varies as a function of strain state. The epitaxial strain significantly affects the $H_{\rm c2}$ slope, since the strain evolves another hole Fermi surface pocket which has a small Fermi energy and large effective mass.\cite{20} Indeed, our Fe(Se,Te) film has a 2$\xi_{\rm c}(0)=2\frac{\xi_{\rm ab}(0)}{\Gamma_{H^{\rm orb.}_{\rm c2}}(0)}\approx$ 0.4\,nm,  which is shorter than the interlayer distance between Fe--Se(Te) layers, $d=0.605\,{\rm nm}$. $H^{||c}_{\text{c2}}(0)$ was estimated to 52.4\,T by WHH model, yielding $\xi_{\rm ab}(0)=\sqrt{\frac{\phi_0}{2\pi H^{||c}_{\text{c2}}(0)}}\approx$ 2.5\,nm, and the $d$ is identical to the out-of-plane lattice parameter, which was calculated using the Nelson--Riley function.\cite{21} Such a short out-of-plane coherence length has been also reported for single crystals.\cite{31}

\begin{figure}
	\centering
		\includegraphics[width=\columnwidth]{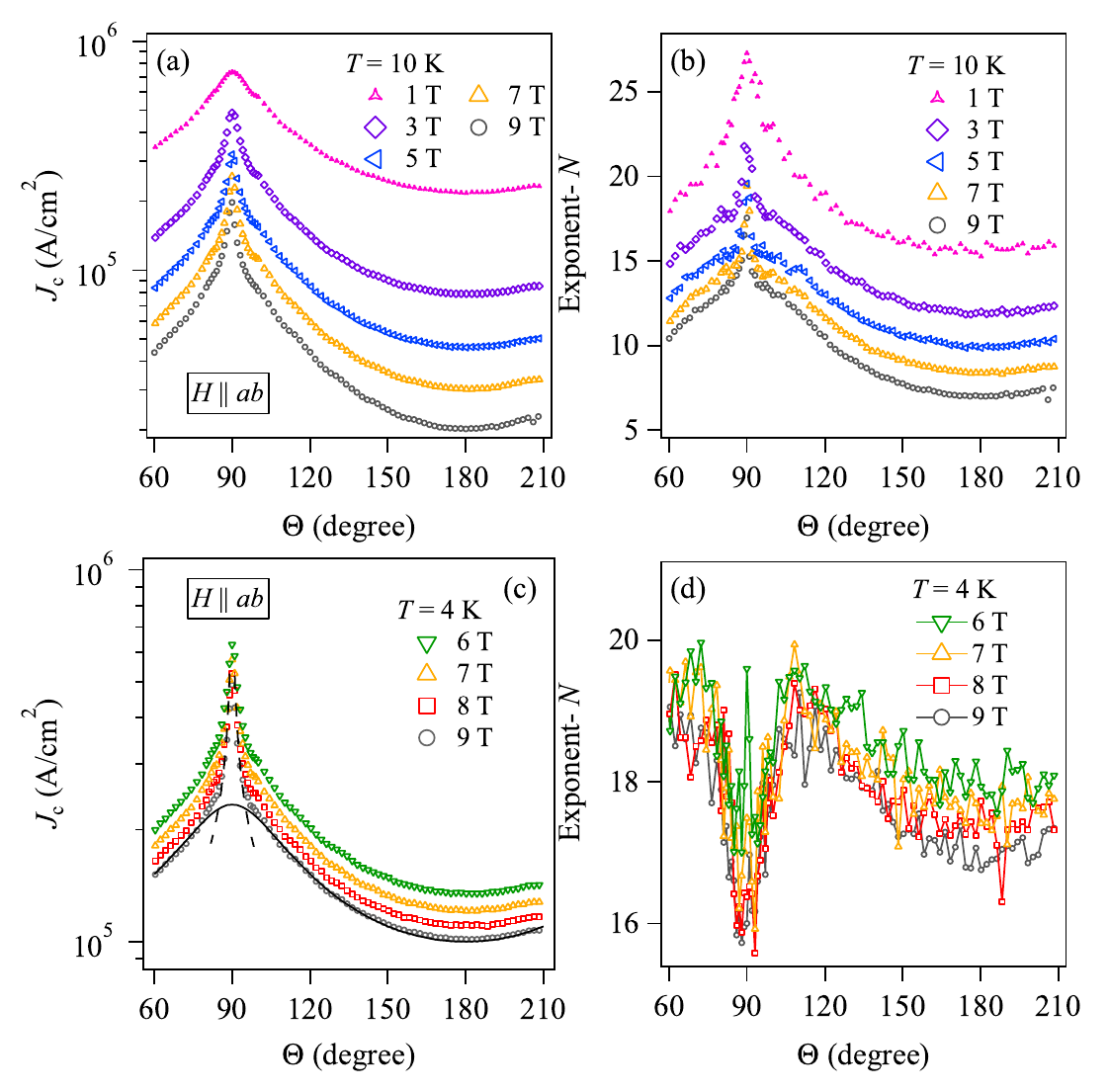}
		\caption{(Color online) (a) Angular dependent $J_{\rm c}$ for the Fe(Se,Te) film and (b) the corresponding $N$ values measured at 10\,K under various magnetic fields. (c) $J_{\rm c}(\Theta$) and (d) the corresponding $N(\Theta$) measured at 4\,K in the range of $6<\mu_0H<9$\,T. The solid and broken lines in (c) represent the random defect and intrinsic contributions at 9\,T, respectively.} 
\label{fig:figure4}
\end{figure}

The $E$-$J$ curves for determining $J_{\rm c}$ show a power-law relation with an exponent $N$, indicative of current limitation by depinning of flux lines rather than grain boundary effects. Angular dependent $J_{\rm c}$ and the corresponding exponent $N$ ($E$$\sim$$J^N$, where $E$ is electric field) measured at 10 and 4\,K for "Bridge\,3" are presented in Figs.\,\ref{fig:figure4}. For both temperatures, $J_{\rm c}(\Theta)$ always has a broad maximum positioned at $\Theta$=90$^{\circ}$ ($H\parallel ab$) which is getting sharper with increasing applied field (Fig.\,\ref{fig:figure4}(a) and (c)). Additionally, no $J_{\rm c}$ peaks at $\Theta$=180$^{\circ}$ were observed in the whole range of temperatures as well as magnetic fields, which is consistent with the TEM microstructural observation shown in Fig.\,\ref{fig:figure2}. Since the exponent $N$ is proportional to the pinning potential, $U_{\rm p}$,\cite{22,23} field-angular dependent critical current density, $J_{\rm c}(H,\Theta)$, curves should be similar to $N(H,\Theta)$. As expected, $N(\Theta)$ behaves almost identically to $J_{\rm c}(\Theta)$ at 10\,K (Fig.\,\ref{fig:figure4}(b)). In contrast, $N(\Theta)$ at 4\,K shows a dip at around $\Theta$=90$^{\circ}$ (Fig.\,\ref{fig:figure4}(d)). Additionally, a tiny peak at $\Theta$=90$^{\circ}$ is observed which develops with decreasing applied magnetic field. Such inverse correlation between $J_{\rm c}(\Theta)$ and $N(\Theta)$ has been observed in YBa$_2$Cu$_3$O$_7$ due to intrinsic pinning, which originates from the modulation of the superconducting order parameter along the $c$-axis.\cite{24,25} The intrinsic pinning contribution to $J_{\rm c}$ can be described by the Tachiki--Takahashi model.\cite{36} As can be seen in Fig.\,\ref{fig:figure4}\,(c), $J_{\rm c}$ close to $H\parallel ab$ can be fitted by this model.

\begin{figure}
	\centering
		\includegraphics[width=\columnwidth]{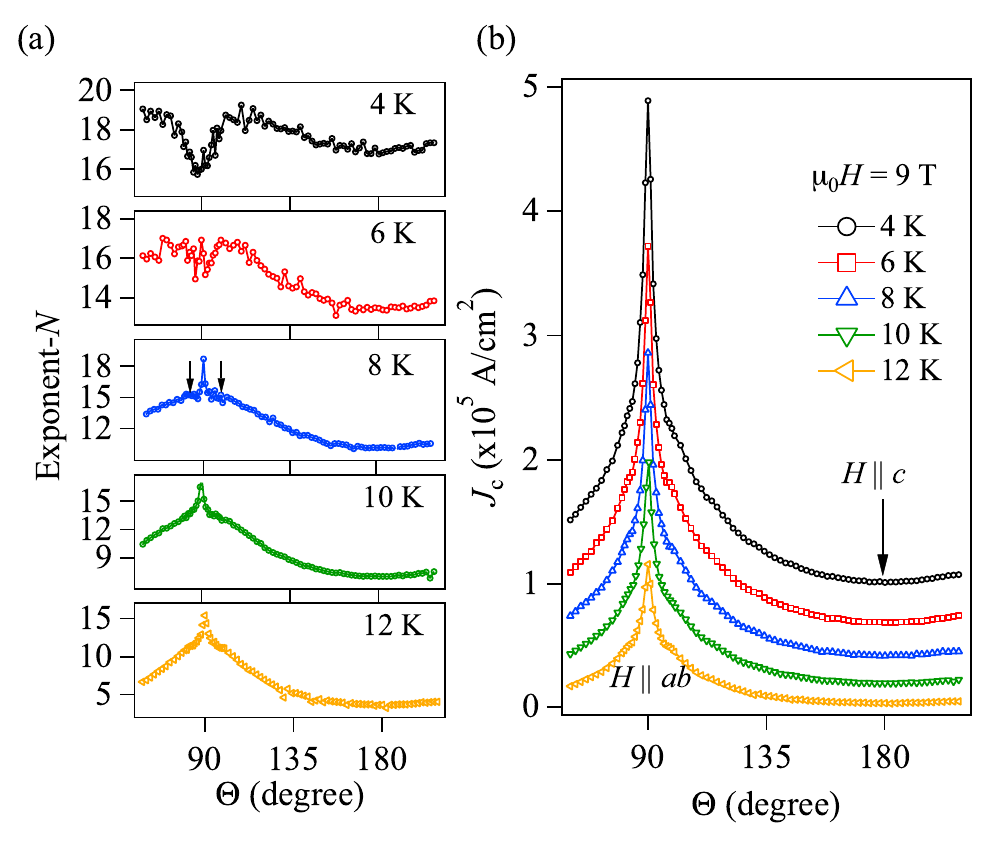}
		\caption{(Color online) (a) Evolution of $N(\Theta)$ as a function of temperature measured at a fixed magnetic field of 9\,T and (b) the corresponding $J_{\rm c}(\Theta$).} 
\label{fig:figure5}
\end{figure}

A dip of $N(\Theta$) is a consequence of  the double-kink excitation of vortices.\cite{26} Blatter $et$ $al$. argued that the activation energy in the stair case regime for intrinsic pinning is increased when the applied field is away from the $ab$-plane.\cite{23} This could explain qualitatively an inverse correlation between $J_{\rm c}(\Theta)$ and $N(\Theta)$.



\begin{figure*}
	\centering
		\includegraphics[width=17cm]{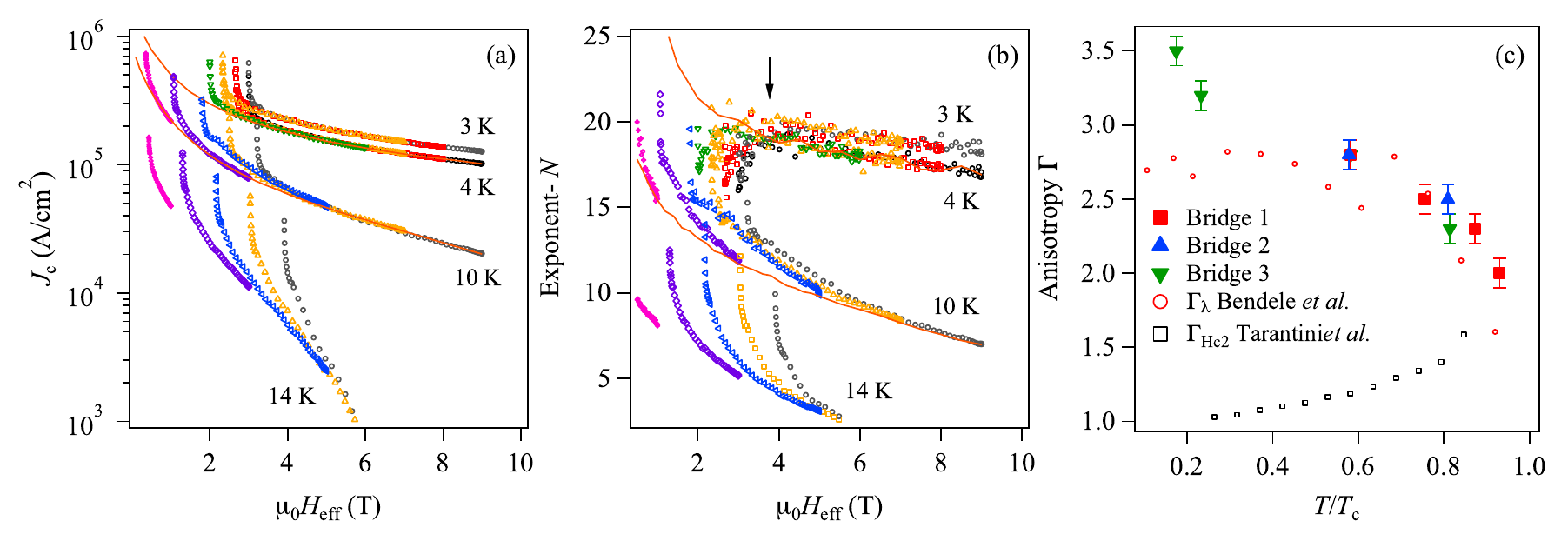}
		\caption{(Color online) (a) The scaling behavior of $J_{\rm c}(\Theta)$ as a function of $H_{\rm eff}$ at various temperatures for "Bridge\,3". (b) The corresponding scaling behavior of $N(\Theta)$. The solid lines represent the measured $J_{\rm c}{\rm (}H\parallel c{\rm )}$ and $N{\rm (}H\parallel c{\rm )}$ at 4 and 10\,K. At 4\,K, the exponent $N$ deviates from the master curve negatively close to $H \parallel ab$ as indicated by the arrow. (c) The $\Gamma$ values obtained by the AGL scaling were observed to increase with decreasing temperature. Here the data from "Bridge\,1 and 2" are also plotted. $\Gamma_{\lambda}$ from $\mu$SR measurements for single crystals by Bendele $et$ $al$.(Ref.\cite{29}) and $\Gamma_{H_{\rm c2}}$ from magneto-transport measurements for strained films by Tarantini $et$ $al$.(Ref.\cite{20}) are plotted for the aim of comparison.}
\label{fig:figure6}
\end{figure*}

We evaluate active transition temperature to intrinsic pinning in our Fe(Se,Te) film by measuring angular dependence of $N$ at various temperature. Since intrinsic pinning is more pronounced in high fields, the maximum field of 9\,T in our experimental condition was employed. Figures\,\ref{fig:figure5} show $N(9\,{\rm T},\Theta)$ and the corresponding $J_{\rm c}(9\,{\rm T},\Theta)$ measured at various temperature. It is clear from Fig.\,\ref{fig:figure5}\,(a) that intrinsic pinning starts being active at a temperature between 10 and 8\,K, since $N$ starts to have shoulders at 8\,K as indicated by the arrows followed by a dip with decreasing $T$. This temperature is almost the same as what we estimate by the following BCS relation, $(1-(\frac{2\xi_{\rm c}(0)}{d})^2)T_{\rm c}=(1-(\frac{0.4}{0.605})^2)\times 17.3\approx 9.7\,{\rm K}$, where 2$\xi_{\rm c}(T)$ is equal to $d$.

All measured $J_{\rm c}{\rm (}\Theta{\rm )}$ are re-plotted as a function of effective field $H_{\rm eff}$, where $H_{\rm eff}$ is the product of $H$ and the scaling function $\epsilon{\rm (}\Theta{\rm )}=\sqrt{\cos^2(\Theta)+\Gamma^{-2}\sin^2(\Theta)}$, where the scaling parameter $\Gamma$ is the mass anisotropy ratio for clean, single-band superconductors.\cite{23} As shown in Fig.\,\ref{fig:figure6}\,(a), all data except for those in the vicinity of $H\parallel ab$ collapse onto the measured curves $J_{\rm c}(H\parallel c$) with $\Gamma$ values of 2$\sim$3.5. Random defects contribution to $J_{\rm c}$ is replotted in the $J_{\rm c}{\rm (}\Theta{\rm )}$ graph at 4\,K and 9\,T (see solid lines in Fig.\,\ref{fig:figure4}\,(c)). Scaling behavior of $N{\rm (}\Theta{\rm )}$ is also shown in Fig.\,\ref{fig:figure6}\,(b). It is apparent that $N{\rm (}\Theta{\rm )}$ can be scaled except for the angular range close to $H \parallel ab$. For $T\leq$4\,K, the $N$ value deviates from the master curve negatively close to $H \parallel ab$ as indicated by the arrow, whereas the opposite deviation is observed above 10\,K, which is due to intrinsic pinning at low temperatures.

In Fig.\,\ref{fig:figure6}\,(c), the extracted temperature dependence of $\Gamma$ is presented. The $\Gamma$ values obtained from different bridges (i.e. Bridge\,1 and 2) are also plotted. Scaling behavior of "Bridge\,1" is presented in Fig.\,\ref{fig:figureS2} in the Appendix A. The scaling parameter $\Gamma$ is observed to increase with decreasing temperature, which is different from what we observed in Co-doped Ba-122 and La-1111.\cite{06,07,08,09} This temperature dependence of $\Gamma$ is similar to $\Gamma_{\lambda}(T)$ rather than $\Gamma_{H_{\rm c2}}(T)$.\cite{29} A similar $J_{\rm c}$ scaling in low field regime, which yields $\Gamma_{\lambda}$, has been reported for MgB$_2$ films.\cite{30} In that case $\Gamma_{\lambda}$ is observed to decrease with decreasing temperature,\cite{37} in contrast to our Fe(Se,Te) film, where it shows the opposite behavior. At low temperatures $\Gamma_{\lambda}$ is almost 1 for MgB$_2$ since the Fermi velocity is almost isotropic. In contrast, $H_{\rm c2}$ is almost isotropic for Fe(Se,Te) at low temperatures.\cite{20}

We compare our Fe(Se,Te) thin films with La-1111, where both systems show weakly 2-dimensional superconductivity. The respective $\frac{\xi_{\rm c}(0)}{d}$ for La-1111 and Fe(Se,Te) are 0.48 and 0.33, indicative of weakly 2-dimensional superconductivity. On the other hand, Co-doped Ba-122 shows 3-dimensional rather than 2-dimensional behavior, since its value of $\frac{\xi_{\rm c}(0)}{d}$ is larger than 1. Our Fe(Se,Te) thin film might be in the clean limit, similarly to the films reported by Tarantini $et$ $al$.\cite{20} They argue that their strained Fe(Se,Te) film is in the Fulde--Ferrel--Larkin--Ovchinnikov state at low $T$ and high $H$, which requires clean limit. On the other hand, our previous La-1111 thin film is in the dirty limit, since the $\xi_{\rm ab}(0)\approx3\,{\rm nm}$ is slightly longer than the Drude mean free path (2.5\,nm). It is noted that the temperature dependence of $\lambda$ and $\xi$ anisotropy for MgB$_2$ strongly depends on its purity.\cite{Golubov,Kogan} In the clean limit, the $\Gamma_{\lambda}$ is observed to decrease with decreasing temperature. On the other hand, the $\Gamma_{\lambda}$ shows weak temperature dependence in the dirty limit. It might be possible that the temperature dependence of $\lambda$ and $\xi$ anisotropy for Fe(Se,Te) or even for other oxypnictides are also similar to that of MgB$_2$ but with opposite behavior. (i.e. $\Gamma_{\lambda}$ is increased with decreasing $T$ in the clean limit for Fe(Se,Te) or oxypnictides, whereas MgB$_2$ behaves in the opposite way.) However, the above discussion is largely speculation. Further investigation is underway.

\section{Conclusions}
We have investigated transport properties of clean epitaxial Fe(Se,Te) thin films prepared on Fe-buffered MgO (001) single crystalline substrates. TEM investigation revealed that the films are free from correlated defects and large angle grain boundaries. Additionally, a sharp interface between Fe(Se,Te) film and Fe buffer has been realized. The $T_{\rm c}$ of the film was 17.3\,K, which is higher than the bulk value, due to compressive strain. A steep slope of 74.1\,T/K in the upper critical field for $H||ab$ was observed, indicating a very short out-of-plane coherence length, $\xi_{\rm c}(0)$, yielding 2$\xi_{\rm c}(0)\approx0.4\,{\rm nm}$. This value is shorter than the interlayer distance between Fe--Se(Te) planes, resulting in modulation of the superconducting order parameter along the $c$-axis and hence intrinsic pinning. These pinning centers are found to be effective below 10\,K. The angular dependent $J_{\rm c}$ as well as the corresponding exponent $N$ can be scaled with the anisotropic Ginzburg Landau theory with a scaling parameter, which follows the penetration depth anisotropy, $\Gamma_{\lambda}$. 

\begin{acknowledgments}
The authors would like to thank M.\,Weigand and B.\,Maiorov of Los Alamos National Laboratory, C.\,Tarantini of National High Magnetic Field Laboratory, as well as S.\,Haindl, V.\,Grinenko, G.\,Fuchs for fruitful discussions, J.\,Scheiter for preparing FIB cuts and TEM lamellae, J.\,Engelmann, M.\,K\"{u}hnel, U.\,Besold for their technical support. The research leading to these results has received funding from European Union's Seventh Framework Programme (FP7/2007-2013) under grant agreement number 283141 (IRON-SEA). This research has been also supported by Strategic International Collaborative Research Program (SICORP), Japan Science and Technology Agency. A part of the work was supported by DFG under Project no. BE\,1749/13. S.\,W.\,acknowledges support by DFG under the Emmy-Noether program (Grant no.\,WU595/3-1).
\end{acknowledgments}

%

\clearpage
\appendix

\setcounter{figure}{0}
\renewcommand{\thefigure}{\thesection\arabic{figure}}

\begin{widetext}
\section{Transport measurements employing various bridges}

Resistance curves as a function of temperature for 3 different bridges and the un-patterned film are summarized in Fig.\,\ref{fig:figureS1}\,(a). For all samples, the resistance curves almost identically vary with temperature. Shown in Fig.\,\ref{fig:figureS1}\,(b) is the normalized $J_{\rm c}{\rm (}H{\rm )}$ curves at 10\,K for the corresponding samples presented in Fig.\,\ref{fig:figureS1}\,(a). The data is normalized by the self-field  $J_{\rm c}$ ($J_{\rm c}^{\rm s.f.}$). $J_{\rm c}^{\rm s.f.}$ values fluctuate with a variation of 30\,\% due to measurement errors of dimensions of the bridges. All bridges behave almost identical. These results prove that the film is homogeneous.

In Fig.\,\ref{fig:figureS2}\,(a), the scaling behavior of $J_{\rm c}(\Theta)$ for "Bridge-1" is displayed. It is clear that all $J_{\rm c}(\Theta)$ curves measured at several temperature can be scaled with $\Gamma$ values of 2$\sim$3. The corresponding exponent-$N$ can be also scaled, as shown in Fig.\,\ref{fig:figureS2}\,(b).

\begin{figure}[h]
	\centering
		\includegraphics[width=10cm]{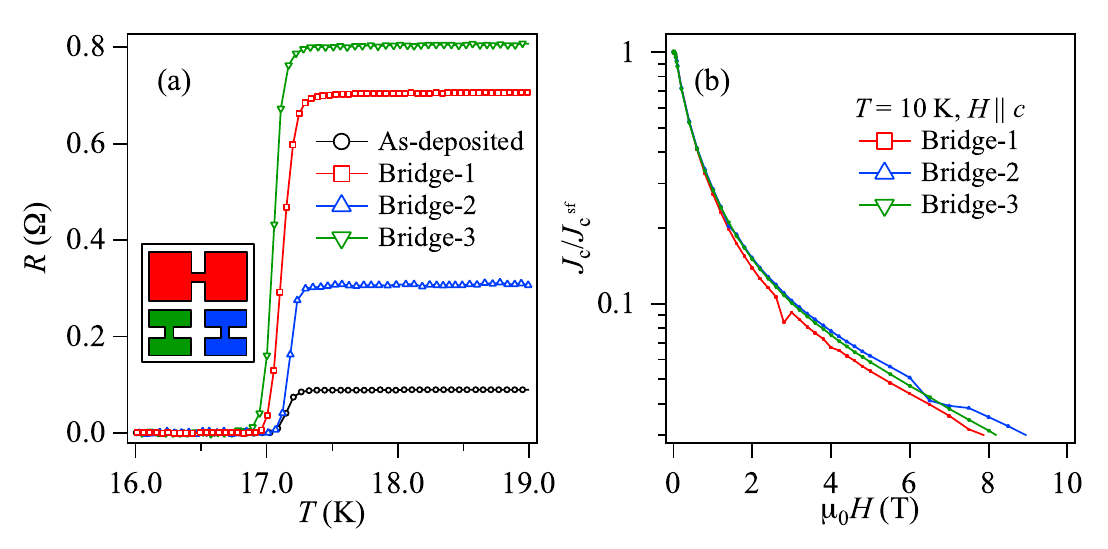}
		\caption{(Color online) (a) Resistance curves of 3 different bridges and un-pattern film. Each bridge is also schematically illustrated. Filled color corresponds to the each trace's one. All samples show the almost identical behavior in $R-T$ with a high $T_{\rm c}$ of around 17\,K. (b) Field dependence of $J_{\rm c}/J_{\rm c}^{\rm s.f.}$ of 3 different bridges measured at 10\,K. Here $J_{\rm c}^{\rm s.f.}$ is self-field $J_{\rm c}$.} 
\label{fig:figureS1}
\end{figure}

\begin{figure}[!bh]
	\centering
		\includegraphics[width=10cm]{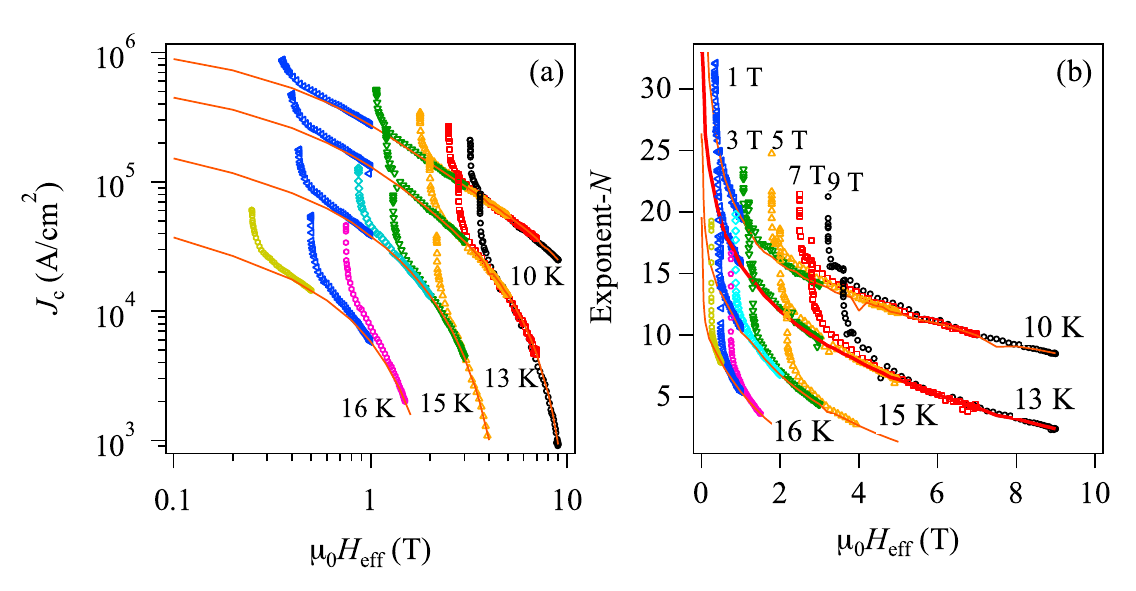}
		\caption{(Color online) (a) The scaling behavior of $J_{\rm c}(\Theta)$ as a function of $H_{\rm eff}$ at various temperatures. The solid line represents the measured $J_{\rm c}{\rm (}H{\rm )}$ for $H \parallel c$. (b) The corresponding scaling behavior of $N(\Theta)$. The solid line represents the measured $N{\rm (}H{\rm )}$ for $H \parallel c$.} 
\label{fig:figureS2}
\end{figure}

\end{widetext}

\end{document}